# Pulse-width modulated oscillations in a nonlinear resonator under two-tone driving as a means for MEMS sensor readout

Samer Houri[1*], Ryuichi Ohta[1], Motoki Asano[1], Yaroslav M. Blanter[2], and Hiroshi Yamaguchi[1*]

[1]*NTT Basic Research Laboratories, NTT Corporation, 3-1 Morinosato-Wakamiya, Atsugi-shi, Kanagawa 243-0198, Japan*

[2]*Kavli Institute of Nanoscience, Delft University of Technology, Lorentzweg 1, 2628 CJ Delft, The Netherlands*

[*]E-mail: Houri.Samer@lab.ntt.co.jp, Yamaguchi.Hiroshi@lab.ntt.co

A MEMS Duffing resonator is driven by two adjacent frequency tones into the nonlinear regime. We show that if the two-tone drive is applied at a frequency where a bistable response of the nonlinear oscillator exists, then the system output will be modulated by a relaxation cycle caused by periodically jumping between the two solution-branches of the bistable response. Although the jumps are caused by the beating of the drives, the existence and period of this relaxation or hysteresis cycle is not solely dictated by the beat frequency between the two driving tones, but also by their amplitude and detuning with respect to the device resonance frequency. We equally demonstrate how the period of the cycles can be tuned via added tension in the device and how these oscillations can be used as a means of sensitive pulse-width modulated (PWM) readout of MEMS sensors.

## 1. Introduction

Micro-electromechanical systems (MEMS) have become ubiquitous as sensors and actuators, their particular advantage being their ability to directly transduce forces, for example pressure or inertial forces, to an electrical output (or vice versa).[1-3] Therefore, it is of continuous scientific and industrial interest to pursue new means of sensor design and/or readout that offer improvement either in terms of sensitivity or implementation simplicity. Equally, MEMS systems are an excellent platform for the experimental investigation of nonlinear dynamics as they offer ease of transduction, pronounced nonlinear parameters, and convenient time scales.[4-5] MEMS devices can therefore be used to demonstrate modes of operations that are otherwise inaccessible with purely linear systems.[6] This work will demonstrate that a MEMS nonlinear behavior can be harnessed to produce a periodic relaxation response that can be used as sensitive MEMS sensor readout.

Relaxation oscillators are a class of self-sustained devices in which a continuous energy discharge forces the system to periodically change states. Relaxation oscillators are characterized by the existence of two distinct timescales, whereby a potential builds in the system according to a slow timescale, followed by a rapid release dictated by a fast time scale.[7] Examples of relaxation oscillators abound in physical and biological systems, such as for instance the Pearson-Anson neon lamp oscillator, micro-electromechanical impact oscillators, nanofluidic devices, quartz crystals, and the mammalian heart muscle.[8-12]

As such, relaxation oscillators are self-sustained oscillators in the sense that they exhibit a stable limit cycle which requires no external periodic drive. Indeed, they are an approximation of the van der Pol oscillator for the case of extremely large negative damping.[13] On the other hand, relaxation-like cycles (also known as hysteresis cycles) that result from an external drive to a hysteretic system are less explored in the context of MEMS devices, and are therefore the focus of this work. The most common example of hysteresis cycles in an electronic system is perhaps the sinusoidal drive of an memristor.[14] In the case of the MEMS device studied here, the slow time scale is provided by a small frequency spacing between two driving tones, while the fast time constant is provided by the system's response. Note that unlike limit cycle relaxation oscillators this system is periodically driven, but exhibits a relaxation-like (hysteresis) cycle as the device is pulled repeatedly through a birfurcation zone.

This work further builds upon previously reported results by investigating the dynamics of these cycles in more detail.[15] In Sect. 2 we describe the MEMS device and the experimental setup used in this work. Section 3 demonstrates the observation of hysteresis cycles for the case when the MEMS nonlinear resonator is subjected to a two-tone drive, we also determine the conditions necessary for such cycles to appear and present a simplified model that qualitatively reproduces the observed dynamics. Section 3.1 explores how the period of the cycles can be modified by the addition of a stimulus (in the form of an added tension to the structure), and demonstrate that they could act as a simplified pulse-width modulation (PWM) readout scheme for MEMS sensors. Section 3.2 demonstrates that the PWM readout can be used as a phase detector for a low frequency signal (with a period on the order of the two-tone beat), a behavior that is equally reproduced using the model. The system shows interesting dynamics that take place at the time scale of the slow oscillations, this is demonstrated in Sect. 3.3 where when two distinct modes are forced to undergo hysteretic oscillations, their cycles exhibit rich interactions.

## 2. Experimental methods

The MEMS resonator used in this work is a freestanding doubly-clamped beam structure 150 μm long, 20 μm wide, and fabricated from an 600 nm thick $Al_{0.3}Ga_{0.7}As$/GaAs heterostructure. The heterostructure's interface forms a two-dimensional electron gas (2DEG) layer that is used as one of two transduction electrodes, and that is Ohmically contacted via a gold-germanium-nickel diffusive contact. The GaAs layer acts as piezoelectric transducer that undergoes mechanical deformation upon the application of a voltage across it, and a top gold layer is deposited to act as a counter-electrode that allows the application of a transducing signal.[16]

The device is wire bonded and placed in a vacuum chamber (chamber pressure ~ $10^{-4}$ Pa) with electrical and optical access. Electrical signals are supplied by WF1974 waveform generators, whereas mechanical motion is detected via an NEOARK laser Doppler vibrometer (LDV) who's output signal is fed into an HP89410A vector signal analyzer (VSA).

To establish the basic device parameters we use a lock-in amplifier (Zurich Instruments HF2LI) to perform a low power (~ 100 $mV_{PP}$) frequency sweep around 1.5 MHz where a resonant mode is detected (corresponding to the structure's fifth mode) as shown in Fig. 1(a). The linear frequency response is fitted to give a resonance frequency of $f_{05}$ = 1.567 MHz, and a quality factor of 1272 corresponding to a line width of 1232 Hz. As the drive voltage is increased to 2 $V_{PP}$ the device exhibits a nonlinear response with a positive Duffing coefficient, i.e. a stiffening nonlinearity, also shown in Fig. 1(a). The frequency is swept forwards and backwards as to measure the hysteresis window, i.e. the region where two stable solutions coexist, as shown in the same figure. The frequency sweep is equally performed for a range of drive amplitudes in order to construct a full map of the location of the hysteresis region as a function of drive and detuning, shown in Fig. 1(b). Note that a constant -2$V_{DC}$ potential is maintained at all times to keep the gold-GaAs Schottky contact in reverse bias when a drive signal is applied.

## 3. Results and discussion

Upon the application of a two frequency components drive, the device responds in a manner that is highly dependent on the amplitude and the detuning of the drive tones with respect to the resonance frequency. For example, if two 1.5 $V_{PP}$, i.e. 1.5 $V_{PP}$ each, drive tones that are 10 Hz apart are applied at a frequency of 1.5778 MHz, then we observe a relatively large harmonic amplitude in the spectral response of the system, whereas if the drive frequency is shifted to 1.5779 MHz, a shift of only 100 Hz, then the frequency response shows a markedly different behavior with significantly reduced harmonic generation, as shown in Fig. 2(a). The 10 Hz spacing between the driving tones is selected to be much smaller than the line width of the resonator so as to allow enough time to observe transient response of the nonlinear system. Note that unless clearly mentioned otherwise, the spacing between drive tones will be 10 Hz in all the experimental results.

Such behavior, i.e. significant generation of harmonics, has been observed in micro- and nanomechanical systems under two- or multi-tone drive, however it has not been fully pursued.[17-18] In particular, previous literature work focused on the spectral response of such systems without looking into the time-domain dynamics. Investigation of the time-domain response using the VSA at time scales corresponding to ~ 10 Hz reveals an equally pronounced difference between the two cases, where the latter drive conditions show a relatively smoothly varying envelope with an oscillation period set by the beat frequency between the two drive tones, i.e. 10 Hz, as shown in Fig. 2(b). For the former drive conditions on the other hand, a pulse like output waveform is observed, with the sudden transitions explaining the increased amplitude of generated harmonics, also shown in Fig. 2(b). Although these pulses have a repetition rate set by the beat frequency, their width (from here on referred to as hysteresis period or $\tau$) is set by the interplay between the drive amplitude, beat frequency, detuning, and the system's nonlinearity.

The existence of these pulses can be understood conceptually by noting that a two-tone drive is equivalent to a single drive with double the amplitude that is being modulated at the beat frequency, thus the drive force periodically varies between 0 and 2$F_{Max}$. If the system is being driven at a frequency where bistability exists, i.e. a frequency where two solution branches coexist as shown in Fig. 1(b), the system starts on the low amplitude branch solution and follows it as the force increases until the lower branch solution is destroyed in a saddle-node bifurcation and the system is forced to jump to the upper branch solution with a corresponding sudden increase in oscillation amplitude observed as a pulse in the VSA. This is demonstrated in Fig. 2(c) where the vibration amplitude is plotted against the envelope of the drive voltage for the two cases. The plots demonstrate the critical role played by the drive conditions on the existence of the hysteresis cycle, since if the drive force or detuning is not enough to pull the system through the second bifurcation limit, a relaxation cycle will not take place.

Therefore it is possible to summarize the experimental conditions necessary to establish a relaxation-like cycle in the response of the MEMS resonator as, a) the drive tones have to be located at a frequency sufficiently detuned from the resonance frequency as to allow bistability to exist, b) the combined amplitude of the drive tones has to be larger than the upper bifurcation point seen in the amplitude-frequency response in Fig. 1(b).

To produce a model that is capable of capturing these dynamics we start with the equation of a nonlinear Duffing resonator driven by two adjacent frequencies:[19]

$$\ddot{X} + \gamma \dot{X} + \omega_0^2 X + \alpha X^3 = F cos(\omega t) + F cos((\omega + \delta)t). \qquad (1)$$

where $X$ signifies the oscillator's motion, $\gamma$ is the damping, $\omega_0$ is the natural frequency, $F$ is the forcing term (note that we consider the two tones to have equal force amplitudes), and $\delta$ is the frequency detuning between the two tones with $\delta = 10$ Hz $<< \gamma$.

By applying the rotating frame approximation we are able to concentrate our attention only on the slowly varying envelope of the vibration amplitude.[20] Since only the case $\delta << \gamma$ is being considered, it is possible to consider the quasi-stationary solutions which reads:

$$\frac{9}{16}\alpha^2 A^6 - 3\alpha \omega_0^2 \Delta A^4 + (4\Delta^2 + \gamma^2)\omega_0^2 A^2 = F\big(1 + cos(\delta t)\big). \qquad (2)$$

where $A$ is the amplitude of the vibration envelope, and $\Delta = \omega - \omega_0$ is the detuning between the drive and the resonance frequencies. Note that as stated in the previous section the forcing term on the right hand side of Eq. (2) periodically changes between 0 and $2F$. An example solution to Eq. (2) is shown in Fig. 2(d), where the detuning $\Delta$ is chosen as to guarantee the existence of two solution branches and the $2F$ forcing term is enough to guarantee crossing the upper bifurcation limit.

It is possible to further simplify the model by considering a rectangular hysteresis response where the system will jump-up once the drive acting upon it crosses an upper threshold ($T_H$) and jumps back down once a lower threshold ($T_L$) is crossed as shown schematically in Fig. 2(d). Changing the DC bias, or applying another stimulus has the effect of modifying the values of the jump-up and jump-down thresholds and thus the hysteresis period. Using this piecewise model, it is possible to reproduce the effect of a stimuli, such as a low frequency signal, has on the hysteresis period of the system.

### 3.1 Pulse width modulation for sensor readout

Dynamic MEMS sensors usually function by measuring the resonance frequency shift due to an externally applied measurand.[21-25] The frequency shift observed in a MEMS resonant sensor is due to a change in either the effective mass, Refs. 20-24, or the effective stiffness of the structure.[26-27]

Since the period of the relaxation-like cycles investigated in this work depends not only on the drive conditions but also on the properties of the MEMS device, any shift to the resonance frequency would equally affect the location of the hysteresis region thus changing the hysteresis period, this dependence offers an interesting prospects to be used as a simplified PWM MEMS sensor readout mechanism.

To demonstrate this effect experimentally the resonance frequency of the GaAs/AlGaAs MEMS needs to be slightly tuned, this is done by changing the DC bias applied across the device (from -2 V to -1 V) and tracking the corresponding resonance frequency shift ($\Delta f$).[6,16] The dependence of resonance frequency on DC bias is plotted in Fig. 3(a), fitting a linear relation to the data gives a slope of 2.7 kHz/V which corresponds to a percentage change in resonance frequency of 0.17%/V.

To investigate the impact of this frequency shift on the hysteresis period, the MEMS device is subjected to a two-tone drive at an excitation frequency $f_{exc}$ = 1.5702 MHz, and a drive amplitude of 1 $V_{PP}$ per tone (chosen as to guarantee the presence of a relaxation cycle). The impact of a change in DC bias on the hysteresis period is shown in the inset of Fig. 3(b), where for a -2 V bias the hysteresis period ($\tau$) is ~ 53 ms, while for a -1.5 V bias the hysteresis period becomes ~ 78 ms. The procedure is repeated for a range of bias values from -2 to -1.4 V (beyond which hysteresis cycles no longer take place), whereby the extracted hysteresis period, plotted in Fig. 3(b), shows a monotonously increasing dependence on bias voltage, which also corresponds to a monotonously increasing dependence on resonance frequency shift. A linear fit to the data gives a hysteresis period slope of ~ 45 ms/V. Thus, a stimulus that can shift the resonance frequency of the MEMS device by 0.17% would result in an easily detectable 100% change (45 ms) to the hysteresis period.

### 3.2 PWM for phase detection

While the previous section demonstrated that a static stimulus (change in DC bias) can be detected via the change in hysteresis period, this section shows how the same scheme can be used to transform an input signal's phase into a PWM output, and uses the piecewise model to reproduce the observed results.

Since the hysteresis period is sensitive to changes in the resonator properties, applying a low frequency signal ($f_{in} \ll f_{exc}$) to the resonator would periodically modulate its tension (due to stretching induced nonlinearity[28]), and thus periodically changes the hysteresis period. If the input signal has the same frequency as the two-tone beat, then $\tau$ changes by a constant amount that depends on the phase difference between the beat frequency and the input signal. On the other hand if the input signal has a different frequency than the two-tone beat, then the change in $\tau$ will take on a periodic, or quasi-periodic, time dependence, and will not be the focus of this work.

Figure 4(a) demonstrates the phase sensitivity of the hysteresis period upon the application of a 0.2 $V_{PP}$ 10 Hz signal (on top of the -2 $V_{DC}$ and the two-tone drive signals that are 1 $V_{PP}$ each), where for a zero phase difference between the low frequency input signal and the two-tone beat (i.e. $\Delta\phi = 0$) $\tau$ takes a value of 48 ms, whereas the period changes to $\tau = 56$ ms, a change of 17%, when $\Delta\phi = \pi$. The full dependence of the hysteresis period on the phase difference is plotted in Fig. 4(b), note that the plot is somewhat asymmetric as a function of $\Delta\phi$ having a maximum around $\Delta\phi = 1.2\pi$.

Equally, we expect the hysteresis period to show a dependence not only on the phase of the input signal but also on its amplitude. This dependence is measured experimentally by sweeping the amplitude and phase of the input signal and extracting the hysteresis period for each value of the input. The relative change in $\tau$ as a function of the input signal amplitude and phase is plotted in Fig. 4(c). As expected low signal amplitudes tend to have a small effect on the hysteresis period, as the signal amplitude becomes on the order ~ 0.1 $V_{PP}$ not only does $\tau$ start to exhibit a clear dependence on the phase of the input signal, but also the maximum of this dependence, which originally is around $\Delta\phi = \pi$, starts to shift right with increasing signal amplitude.

By using parameters obtained from Fig. 1(b) and Fig. 3(a), i.e. dependence of thresholds on detuning and the dependence of resonance frequency on DC bias, and applying them to the piecewise model described in Sect. 2, it is possible to qualitatively reproduce the phase dependence observed experimentally including the asymmetry that develops at relatively large input signal amplitudes, shown in Fig. 3(d). This asymmetry originate from the fact that $T_H$ and $T_L$ thresholds are affected differently by detuning, and thus by frequency shifts, which is evident by looking at the dependence of the upper and lower thresholds on detuning as observed in Fig. 1(b).

**3.3 Mode-coupling mediated interaction of two relaxation-like cycles**

A particularly interesting aspect of the dynamics of the classical limit-cycle relaxation oscillators is their distinctive ability to form highly interacting oscillator networks.[29-30] These networks have been demonstrated in physical systems, and they are equally considered a canonical model for the interaction of neurons in a network, where the neurons themselves are modeled as relaxation oscillators known as the FitzHugh-Nagumo Oscillator.[31-32] A main reason for this interest in coupled relaxation limit-cycle oscillators is their ability to interact via a "Fast Threshold Modulation" mechanism that makes them able to synchronize much faster than their counterparts, the phase oscillators.[33-34]

As stated previously, the relaxation-like cycles observed in this work are not identical to the canonical limit-cycle relaxation oscillators, nevertheless they show significant similarities. In particular, the effect of modulating the upper and lower thresholds observed in this work is similar to the "Fast Threshold Modulation" mechanism observed in limit cycle relaxation oscillators. It is therefore interesting to investigate whether the interaction of two of these pulse-width modulated oscillations results in an equally rich dynamics as the one observed for the canonical relaxation oscillators.

Rather than couple two physically distinct devices, we create two relaxation-like cycles that are separated in frequency space, since they are excited around the third and fifth (the one described above) resonance modes respectively. The third mode has a resonance frequency of $f_{03}$ = 959 kHz, a quality factor of 1362 (line width of 704 Hz), and exhibits a positive (stiffening) Duffing nonlinearity similar to the fifth mode. Each mode when vibrating adds a slight average tension to the structure, the added tension shifts the frequency (and thus the $T_H$ and $T_L$ thresholds) of the other mode, an effect known as mode coupling that is already explored in literature.[35-37] Thus mode coupling offers a mechanism by which the two oscillation cycles can interact.

Both modes are subjected to two-tone excitations having a 10 Hz separation (with the two-tone beats having the same phase), with excitation frequencies of 962.2 kHz, and 1.568 MHz, and 0.6 and 1 $V_{PP}$ drive amplitudes (per tone) for the 3$^{rd}$ and 5$^{th}$ modes respectively. These drive conditions are selected such that taken separately each dual tone excitation would establish a relaxation oscillation in the respective mode, Fig. 5(a).

However, as the excitations are applied to both modes simultaneously, the behavior changes drastically and the relaxation oscillations take on a relation in which the oscillations of the 3$^{rd}$ mode (that happens to be of higher amplitude) seem to suppress that of the 5$^{th}$ mode, as can be seen in Fig. 5(b). The qualitative change in behavior of these cycles is a clear indication of a mode-coupling mediated interaction between the two frequency-spaced oscillations, the dynamics of which are yet to be fully understood.

## 4. Conclusions

In summary, this paper presented the dynamics of a nonlinear MEMS resonator when driven by two adjacent frequencies. It was shown that under adequate drive conditions relaxation-like or hysteresis cycles emerge, where the vibration envelope shows a pulse-like response. The cycles are due to jumps between a high and a low amplitude solution branches as the system is periodically pulled through bifurcation points.

It was equally demonstrated that the width of pulses can act as an efficient PWM readout mechanism for any stimulus that can shift the resonance frequency of the structure, even if slightly. The pulse-width was further used as a means to detect the phase of an input signal whose frequency is equal to that of the beat frequency of the two-tone drive, and these results were reproduced using a simplified piecewise model. Finally, this work demonstrated the possibility that multiple relaxation-like cycles that are spaced in the frequency domain can interact with each other, thus preluding the potential for far more complex dynamics.

In addition to signaling the onset of interesting dynamics, the high frequency pulse-width modulated output of a MEMS resonator/sensor can be easily interfaced through an envelope detector to a digital circuit, thus providing a practical low power PWM direct sensor readout.

## Acknowledgments

This work is partly supported by a MEXT Grant-in-Aid for Scientific Research on Innovative Areas "Science of hybrid quantum systems" (Grant No. JP15H05869 and JP15K21727).

## Appendix

# Figure Captions

**Fig. 1.** (Color online) Frequency response plot (in units of $\mu$V/V) of the MEMS resonator in the linear case with $V_{Drive}$ = 100 mV$_{PP}$ (black trace), and the nonlinear response for large amplitude drive $V_{DRIVE}$ = 2 V$_{PP}$ for forwards (blue) and backwards (red) frequency sweeps (a). A two dimensional map of the hysteresis in the resonator's response ($\Delta$H = H$_{forwards}$ - H$_{backwards}$) plotted as a function of $V_{Drive}$ and frequency (b).

**Fig. 2.** (Color online) Observed spectral domain response of a vibrating MEMS beam under two-tone driving for $f_{exc}$ = 1.5778 MHz, red trace, and 1.5779 MHz, blue trace, the inset shows the 10 Hz spacing of the harmonics (a). (b) Time domain response for the same drive conditions, showing the sharp transitions that correspond to large harmonic components in the frequency domain (red trace), also showing the envelope of the driving tones (green). (c) Plot of Amplitude versus Force, showing the origin of the sharp transitions to be jumping between solution branches of the nonlinear oscillator (red trace), the lack of sharp transitions is also indicative of absence of jumps between solution branches (blue trace). (d) Calculated traces from the model showing the effect of the choice of detuning and/or driving amplitude on whether multiple branches exist (red trace) or not (blue trace), also shown is the piecewise simplified model with the two adjustable parameters T$_H$ and T$_L$ (black trace).

**Fig. 3.** (Color online) Measured resonance frequency shift as a function of DC bias voltage (a). (b) Plot showing the dependence of the hysteresis period on $V_{DC}$ and the linear fit (dashed red line). Inset, time domain trace showing the effect of change in bias voltage of the hysteresis period $\tau$, for $V_{DC}$ = -2 and -1.5V (blue and red traces respectively).

**Fig. 4.** (Color online) Time trace of a hysteresis period upon the application of $V_{signal}$ = 0.2 V$_{PP}$ for $\Delta\phi$ = 0 and $\Delta\phi$ = $\pi$ blue and red traces respectively, the effect of the signal phase on the hysteresis period is clearly visible (a). Plot of the dependence of the experimentally

measured hysteresis period ($\tau$) on the phase of input signal, showing the asymmetric nature of this dependence, with the maximum taking place for $\Delta\phi = 1.2\pi$ (b). Dependence of the hysteresis period on the amplitude and phase of the input signal, $\tau$ starts to show significant change when $V_{signal} > 0.1$ $V_{PP}$ (c). Simulations based on the piecewise model of the effect of signal phase on $\tau$, the plot equally reproduces the experimentally observed asymmetry (d).

**Fig. 5.** (Color online) drive conditions that independently establish relaxation cycles for the third and fifth modes, red and blue traces respectively (a). When both cycles are excited simultaneously, mode-coupling mediated interaction results in the fifth mode relaxation cycles being suppressed by the high amplitude vibrations of the third mode (b).

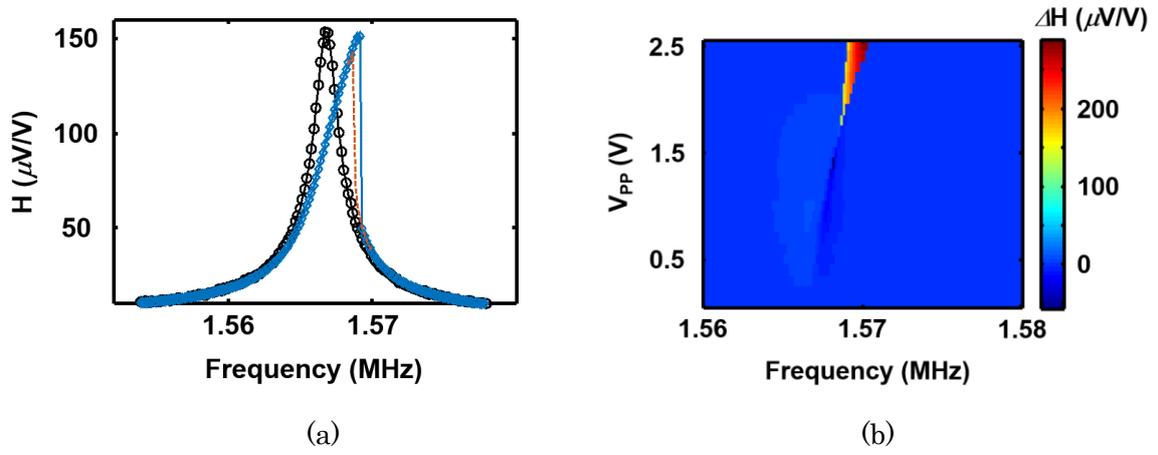

(a)                  (b)

Fig.1. (Color Online)

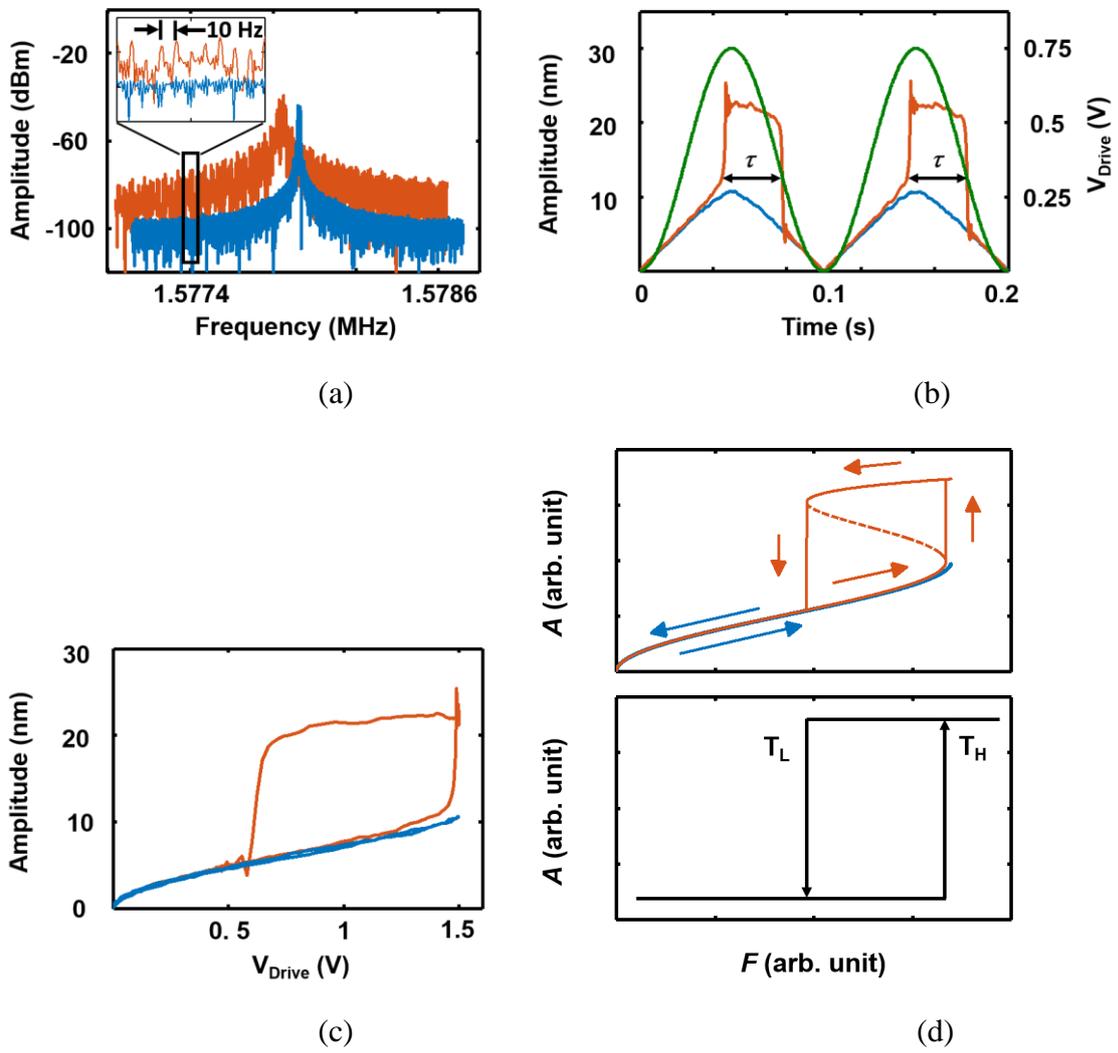

Fig. 2. (Color online).

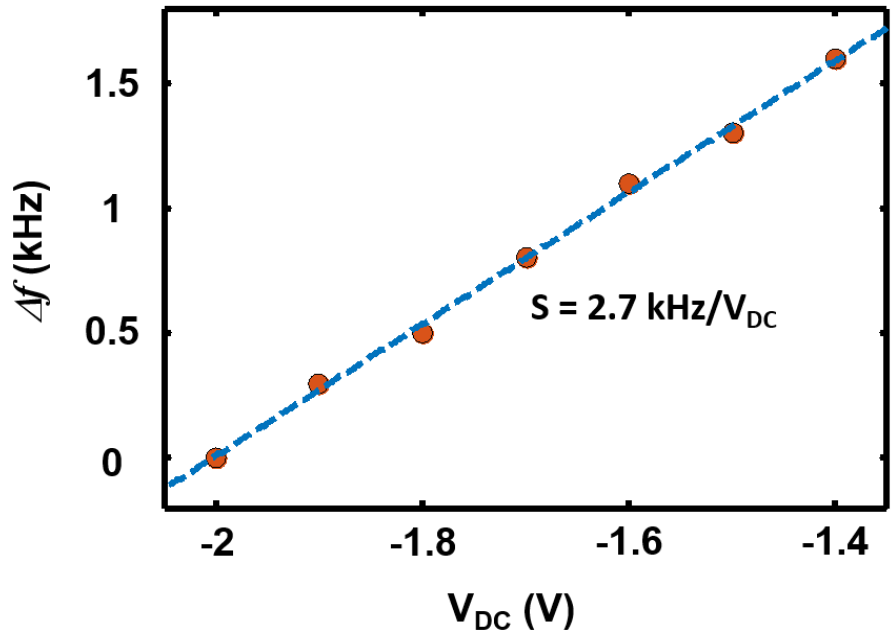

(a)

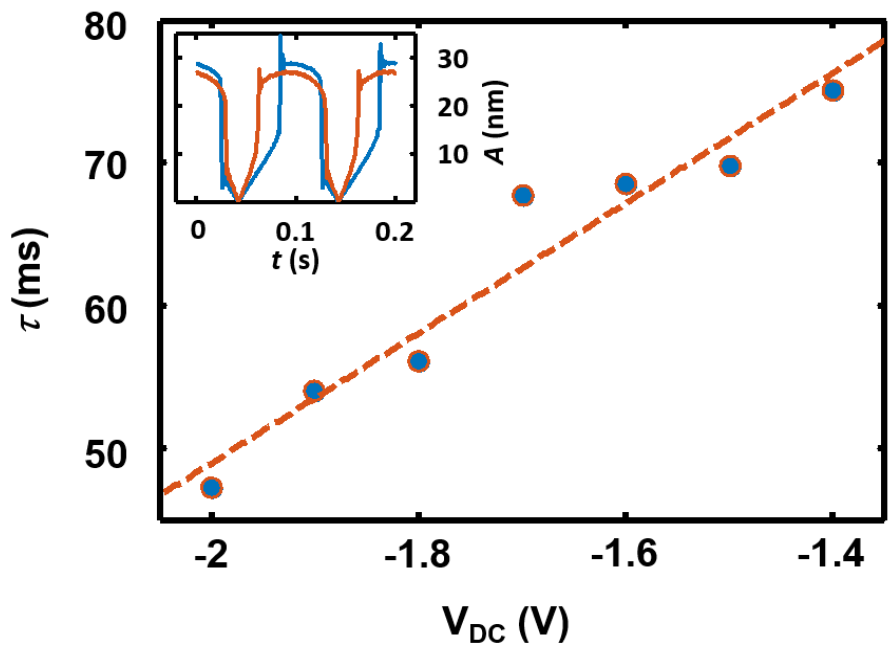

(b)

Fig. 3. (Color online)

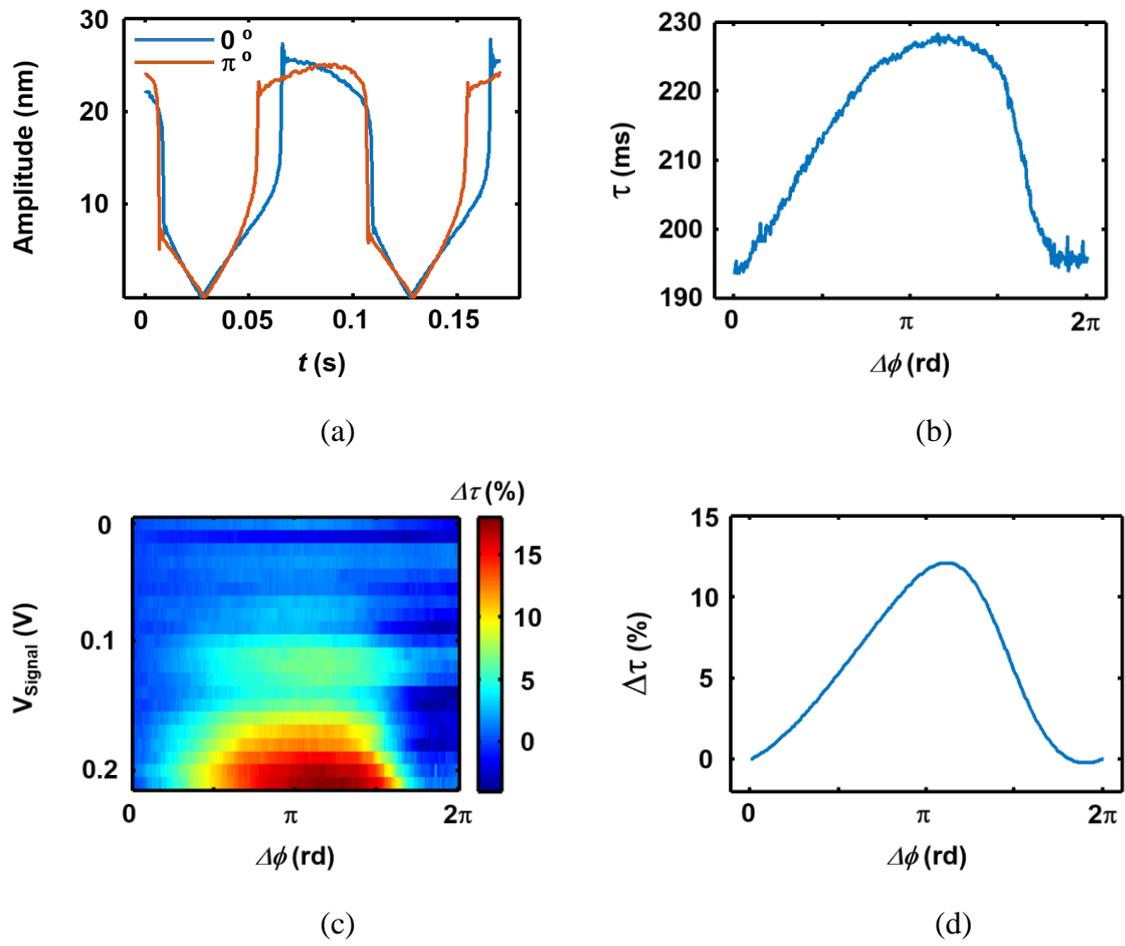

Fig. 4. (Color online)

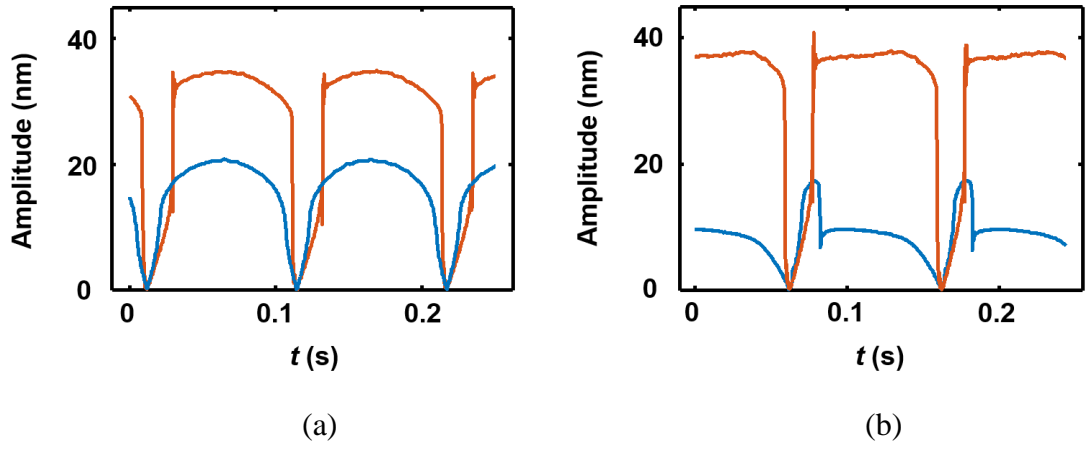

(a)                          (b)

Fig. 5. (Color online)